\documentclass[10pt, conference]{IEEEtran}
\usepackage[utf8]{inputenc}

\usepackage{amsmath,amssymb,amsfonts}
\usepackage{graphicx}
\usepackage{textcomp}
\usepackage{pifont}
\usepackage[bookmarks=false]{hyperref}

\usepackage{amsthm}
\usepackage{xcolor}
\usepackage[figuresright]{rotating}

\usepackage{graphicx}
\graphicspath{{/}{fig/}}

\usepackage{array}
\usepackage{textcomp}
\usepackage{xcolor}
\usepackage{multirow}
\usepackage{booktabs}

\usepackage{mathtools}
\usepackage{float}

\usepackage{pgfplots}
\pgfplotsset{compat=1.7}
\usepgfplotslibrary{groupplots}

\usepackage{multirow,tabularx}
\usepackage{flushend}

\newlength\figureheight
\newlength\figurewidth
\title{\LARGE \bf
    Enhancing Autonomy with Blockchain and \\ Multi-Access Edge Computing in Distributed Robotic Systems
}

\author{
    \IEEEauthorblockN{
        Jorge Peña Queralta\IEEEauthorrefmark{1},
        Li Qingqing\IEEEauthorrefmark{1}.
        Zhuo Zou\IEEEauthorrefmark{2} and
        Tomi Westerlund\IEEEauthorrefmark{1}
    }\\
    \IEEEauthorblockA{
        \IEEEauthorrefmark{1} \href{https://tiers.utu.fi}{Turku Intelligent Embedded and Robotic Systems Lab, Faculty of Sciece and Engineering, University of Turku, Finland} \\
        \IEEEauthorrefmark{2} School of Information Science and Technology, Fudan University, China\\
        Emails: \{jopequ, qingqli, tovewe\}@utu.fi, zhuo@fudan.edu.cn
    }
}

\begin{document}

\maketitle
\thispagestyle{empty}
\pagestyle{empty}

\global\csname @topnum\endcsname 0
\global\csname @botnum\endcsname 0
\begin{abstract}

    This conceptual paper discusses how different aspects involving the autonomous operation of robots and vehicles will change when they have access to next-generation mobile networks. 5G and beyond connectivity is bringing together a myriad of technologies and industries under its umbrella. High-bandwidth, low-latency edge computing services through network slicing have the potential to support novel application scenarios in different domains including robotics, autonomous vehicles, and the Internet of Things. In particular, multi-tenant applications at the edge of the network will boost the development of autonomous robots and vehicles offering computational resources and intelligence through reliable offloading services. The integration of more distributed network architectures with distributed robotic systems can increase the degree of intelligence and level of autonomy of connected units. We argue that the last piece to put together a services framework with third-party integration will be next-generation low-latency blockchain networks. Blockchains will enable a transparent and secure way of providing services and managing resources at the Multi-Access Edge Computing (MEC) layer.
    We overview the state-of-the-art in MEC slicing, distributed robotic systems and blockchain technology to define a framework for services the MEC layer that will enhance the autonomous operations of connected robots and vehicles. 
    
    

\end{abstract}


\IEEEpeerreviewmaketitle

\section{Introduction}

5G and beyond connectivity has the potential for bringing together the telecommunications, robotics~\cite{dohler2017internet}, artificial intelligence (AI)~\cite{li2017intelligent}, Internet of Things (IoT)~\cite{palattella2016internet} and blockchain domains~\cite{backman2017blockchain}, all of which share a recent trend in which computation is shifting towards more distributed architectures~\cite{voigtlander20175g, queralta2020blockchain, wang2019edge}. This comes together with the concept of network slicing and edge computing, key pillars behind the low-latency and network load optimization in 5G and beyond networks~\cite{addad2018towards}. Through multi-tenant slicing, new business opportunities are being created at the edge of the network~\cite{husain2018mobile}. In this study, we provide a vision for the future of 5G-connected robots and vehicles, which will potentially benefit from this connectivity to increase their degree of autonomy and level of intelligence through services offered by third parties at the MEC layer.

We explore the potential for combining the backbone of today's autonomous robotic navigation and localization, the Robot Operating System~\cite{quigley2009ros}, with the latest development in 5G and slicing strategies at the MEC layer. The MEC layer is an inherently distributed computing platform that enables high-performance computing (HPC) services with minimal latency~\cite{sabella2016mobile}. The most direct application is to extend existing offloading schemes~\cite{qingqing2019monocular}, and integrate them within the 5G stack~\cite{zhang2016energy}. This has clear potential in vehicular and robotic navigation, especially when combined with predictive schemes~\cite{zhang2017mobile}. In addition, we envision that FPGA-based and CGRA-based hardware accelerators at the base stations will provide new levels of reconfigurability, energy efficiency, and processing power within the offloading orchestrators. Moreover, we take into account integration between distributed robotic systems~\cite{bareis2019robots}, and distributed computation platforms defined within a blockchain~\cite{wood2014ethereum}. We argue that permissioned blockchains backed by a large public and trusted infrastructure will be a key element of MEC-based services. These blockchains will be able to provide a transparent and secure channel for connected vehicles to interact with third parties.

Slicing at the MEC layer in 5G and beyond can reduce the computational load in connected robots and vehicles. This will allow units with more constrained resources, such as delivery drones, to enhance their situational awareness and increase their autonomy. In terms of safety and reliability in long-term autonomous operation in both self-driving vehicles and autonomous robots, challenges arise from the point of view of (1) localization accuracy \cite{seif2016autonomous}, (2) situational awareness and level of understanding of the environment \cite{hong2019rules}, and (3) limitations of computational capabilities in smaller robots or drones, with algorithms that might take longer to run depending on the complexity of the environment \cite{floreano2015science}. Slicing at the MEC layer has potential for providing services to support the operation of connected autonomous vehicles and robots by providing in respect to the above challenges (1) streaming services of high definition (HD) maps for accurate localization with online updates whenever the environment changes; (2) semantic information of the environment, as well as metadata from other connected vehicles; and (3) an adaptive algorithm that autonomously provides in real-time map models and environment data according to the operational and computational capabilities of the vehicle requesting data.

\subsection{Blockchain at the MEC Layer}

The integration of Blockchain with slicing at the MEC layer has recently been proposed by different researchers \cite{xiong2018mobile, rahman2018blockchain, dai2019blockchain}. Nevertheless, we focus on the use of Blockchain to enable services at the MEC layer for autonomous robots and vehicles. Rather than focusing on data integrity and security, we see that distributed consensus mechanisms in a blockchain are ideal for managing the MEC hardware, services and client-provider interaction. At the same time, the utilization of blockchains in the robotics field has recently shown its potential~\cite{queralta2020blockchain}. We extend it as a framework for integrating external services into distributed robotic and vehicular systems. One of the key challenges in the utilization of a blockchain for applications requiring real-time communication and data processing is the scalability, as indicated by~\cite{ferrer2018blockchain, strobel2018managing}. However, a more wider adoption of blockchain technology across multiple fields, specially the robotics and automation field, is expected with the arrival of low-latency and high-throughput next-generation blockchain networks~\cite{queralta2020blockchain}.

\subsection{Supporting Autonomy in Smart Cities}

\begin{figure}
    \centering
    \includegraphics[width=0.48\textwidth]{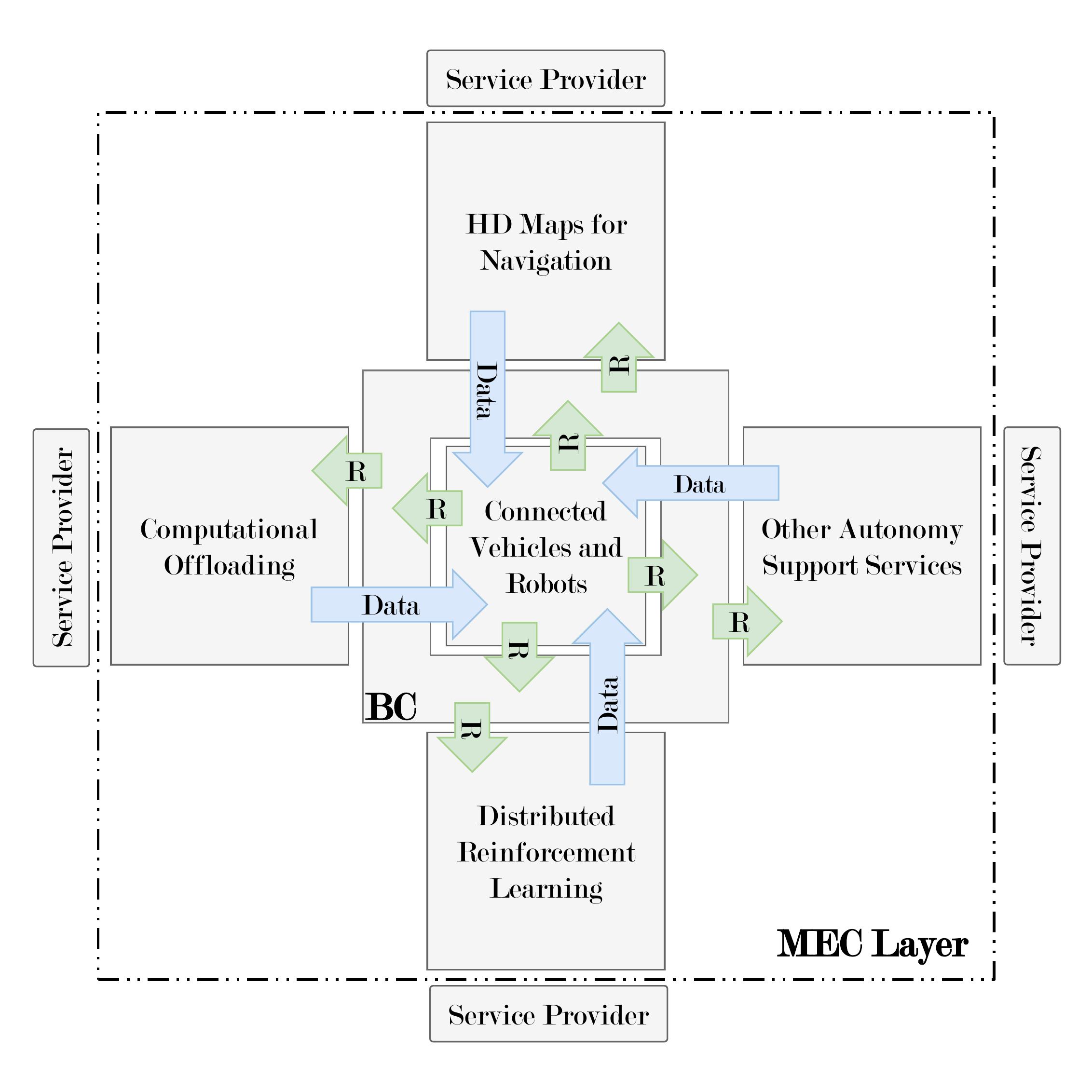}
    \caption{Autonomy Support Services at the Multi-Access Edge Computing (MEC) layer. Requests (small yellow arrows) go through the blockchain slice (BC), which is in charge of the local service orchestration. Data is streamed directly to end-users to reduce latency and increase throughput.}
    \label{fig:services}
\end{figure}

The concept of Smart City has been mostly tied to the IoT since its inception~\cite{smart_city_applications}. Nonetheless, the IoT and the robotics domain have since been integrated as connected robots become the standard. The new edge and fog computing paradigms have only increased this synergy between the two domains~\cite{queralta2019collaborative}. Nam \textit{et al.} surveyed the early works on the topic and defined three fundamental dimensions of a Smart City: technological, human and institutional~\cite{concptualizing_smart_cities}. Self-driving cars or autonomous delivery robots extend the original concept of smart city from passive technology (mainly sensors) to active participators in an increasingly more complex cyber-physical dimension. Nonetheless, little attention has been put on the role of the institutional dimension towards a more widespread penetration of autonomous robots and vehicles in Smart Cities. In a recent work, we argue that institutions and public infrastructure can play a key role in enabling collaboration between autonomous robots with a blockchain~\cite{queralta2020blockchain}. We extend this idea with the introduction of MEC slicing. The key areas of the system are illustrated in Figure~\ref{fig:services}.

Since the introduction of MEC, network slicing has been seen as a key enabler of future autonomous vehicles~\cite{taleb2017multi}. However, the definition of architectures and the specifications of slices have been made mostly from the point of view of the telecommunications domain, taking into account network requirements~\cite{campolo2017slicing}. In a recent work, Mei \textit{et al.} have proposed an intelligent network slicing framework with differentiated slices for (i) traffic safety, (ii) autonomous driving, (iii) infotainment and vehicular internet, and (iv) service slice that manages the previous ones and measures quality of service~\cite{mei2019intelligent}. We believe that efficient slicing requires different granularity within the autonomous driving concept, and therefore, we propose a slicing architecture that takes into account both the network requirements and the type of computing resources utilized for different aspects of autonomous driving, with a clear differentiation between offloading services and streaming services. In addition, we replace the top managing slice in~\cite{mei2019intelligent} with a blockchain-based orchestrating slice that provides a framework for interaction between clients and service providers. Finally, by separating the service orchestration from the data channels and the actual services, which are hosted in their corresponding slice, we argue that the robustness of the system can be increased, where a failure in any of these slices does not have a significant impact on the performance of the others in the short term. To the best of the authors' knowledge, this is the first work to focus on the point of view of algorithmic requirements for slicing, in comparison to a more traditional focus on network requirements. Moreover, we discuss on the key role that public infrastructure can play in ensuring the fairness and transparency of MEC-based services.

The rest of this paper is organized as follows. In Section II, we introduce the basics of MEC, network slicing, permissioned blockchains and algorithms for robotic navigation. Section III overviews the opportunities for utilizing the Blockchain technology, as a framework to manage services and resources with MEC slicing, together with distributed robotic systems. In Section IV, we explore the potential applications of the proposed approach in Smart Cities, from the streaming of local high-definition maps enabling accurate localization to offloading services to enhance the capabilities of robots with more limited resources. Section V then discusses the challenges and opportunities of the proposed architecture, with an emphasis on the viability and scalability of integrating blockchain technology. Finally, Section VI concludes the work and outlines future research directions.

\begin{figure*}
    \centering
    \includegraphics[width=0.95\textwidth]{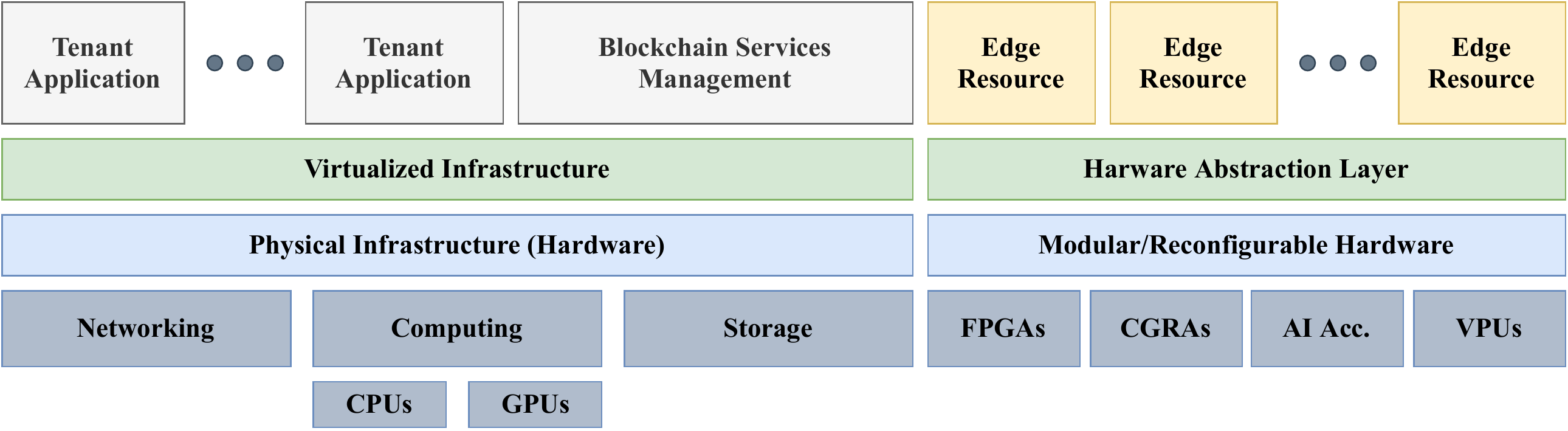}
    \caption{5G-MEC Computational Building Blocks}
    \label{fig:5Garch}
\end{figure*}

\section{Background and Significance}
In this section, we briefly describe the main technologies that are considered in this study: distributed computing platforms at the MEC layer and blockchain technology.

\subsection{Multi-Access Edge Computing and Network Slicing}

The standardization of Multi-Access Edge Computing (MEC) has been promoted by the European Telecommunications Standards Institute (ETSI)~\cite{etsi2018mec}, with the MEC Industry Specification Group (ISG) launched at the end of 2014. The ETSI MEC ISG aims at defining a multi-tenant distributed cloud platform to be located at the edge of the radio access network (RAN)~\cite{hu2015mobile}. Moving computation and data intensive tasks towards the edge of the network enables the low-latency and high-bandwidth requirements of 5G and beyond connectivity. Other fundamental technologies towards this end include containerization and virtualization, software defined networking (SDN), and network function virtualization~\cite{taleb2017multi}.

One of the key pillars enabling multiple verticals within MEC, and opening the RAN edge to a wide variety of industries and users, is network slicing~\cite{nextgen2016study}. Network slicing consists of the co-existence of multiple logical software-defined networks (slices) on a common hardware infrastructure, i.e., a multi-tenant cloud infrastructure at the edge of the network, with each of the slices being optimized to meet the requirements of a particular application~\cite{alliance2016description}. We are particularly interested in slicing for the automotive sector, where 5G will be the key in vehicle-to-everything communication~\cite{giust2018multi}.

Since its early developments, MEC monetization has been a central topic of discussion~\cite{taleb2017multi}. Blockchain can provide a framework to democratize the monetization and utilization of the MEC layer as a platform to offer automation services to connected vehicles or robots. 

From the point of view of security, a recent report from the European Union Agency for Cybersecurity (ENISA) on the thread landscape for 5G networks has identified numerous threats~\cite{enisa}. A blockchain can directly provide a higher level of resilience against multiple of these, such as authentication traffic spikes, manipulation of network traffic, malicious diversion of traffic, among others. Our focus is on utilizing a blockchain as a transparent and distributed framework to achieve consensus in terms of MEC resource provisioning. This has a direct impact on preventing threats such as abuse of third party hosted network functions, manipulation of the network resources orchestrator, or opportunistic and fraudulent usages of shared resources, among others.

In summary, MEC offers the benefits of cloud computing at the edge of the network, with the potential to offer new customer experiences. MEC allows for more scalable applications and network infrastructure by ensuring that raw data is processed at the edge, and only the resulting metadata is transmitted over to central cloud servers or other clients, optimizing the network load and reducing unnecessary traffic with information that does not need to be stored.




\subsection{Blockchain and Distributed Ledgers}

Blockchain platforms can be classified into two main types, permissionless and permissioned~\cite{vukolic2017rethinking}. These can also be denominated public (permissionless), and consortium or private blockchains (permissioned). In public blockchains, there is no authority and all nodes are equivalent. In consortium or private blockchains, there are trusted authorities or nodes in charge of validating transactions \cite{zheng2017overview}. 

One of the most famous and successful blockchains to date is Hyperledger, a project initiated in 2016 within the Linux Foundation \cite{cachin2016architecture}. Hyperledger is a permissioned blockchain which has been successfully utilized in multiple industrial domains \cite{androulaki2018hyperledger}. The objective of Hyperledger is the deployment of an open-source and cross-industry framework that can be utilized as a standard platform to run smart contracts within a decentralized ledger.

Consortium blockchains, and Hyperledger in particular, have key advantages that have an impact over business networks such as a MEC-based blockchain~\cite{ibm2018hyperledger}: (i) all participants have known identities, and therefore data protection laws can be applied accordingly, with permission required in order to join the network; (ii) data partitioning through channels, ensuring that data is available on a need-to-know basis and is only transmitted to the parties that need; (iii) scalability and adaptable levels of trust, with endorsement policies defining the number and nature of validators required to verify a given transaction, and subsequent network load optimization; (iv) a modular architecture, where identities and other components can be easily extended to meet the various requirements of third parties or public authorities; and (v) flexible and complex queries over the ledger, simplifying the auditing process. 

The data partitioning scheme seamlessly combines with the multi-tenancy at the MEC layer, with different service providers not being necessarily aware of other's transactions and data, even if this information is encrypted, as it would happen in a public blockchain. The endorsement policies can be optimized to manage scalability in large networks, defining validators depending on the client's location or the current network load. Then, alternating nodes across the network could be validated in different locations at different times.

The smart contracts within Hyperledger can be leveraged to manage the interaction between clients and service providers. First, by validating transactions and services before data is actually exchanged. Second, by provisioning and reconfiguring the computing resources within the MEC layer that are dedicated to that given service, optimizing the servers' load to improve users' experience. The data itself does not necessarily go through the blockchain once the service has been approved.




We see the main opportunities as part of smart cities, where the blockchain can be supported by either RAN or 5G-connected public infrastructure, as well as industrial environments where there exists trust. Figure~\ref{fig:5Garch} illustrates a generic 5G-MEC architecture with a blockchain to manage services (tenant applications) and edge resources (reconfigurable and on-demand hardware). In a smart city, a public blockchain for data sharing could boost the deployment of autonomous robots from both private and public entities. Besides, the role of the infrastructure should be considered not only as a platform to manage the blockchain lifecycle but also as a static data source and validating platform, where traffic cameras and other sensors that already exist can be integrated.

\section{Blockchain-based services at the MEC layer}

\begin{figure*}
    \centering
    \includegraphics[width=\textwidth]{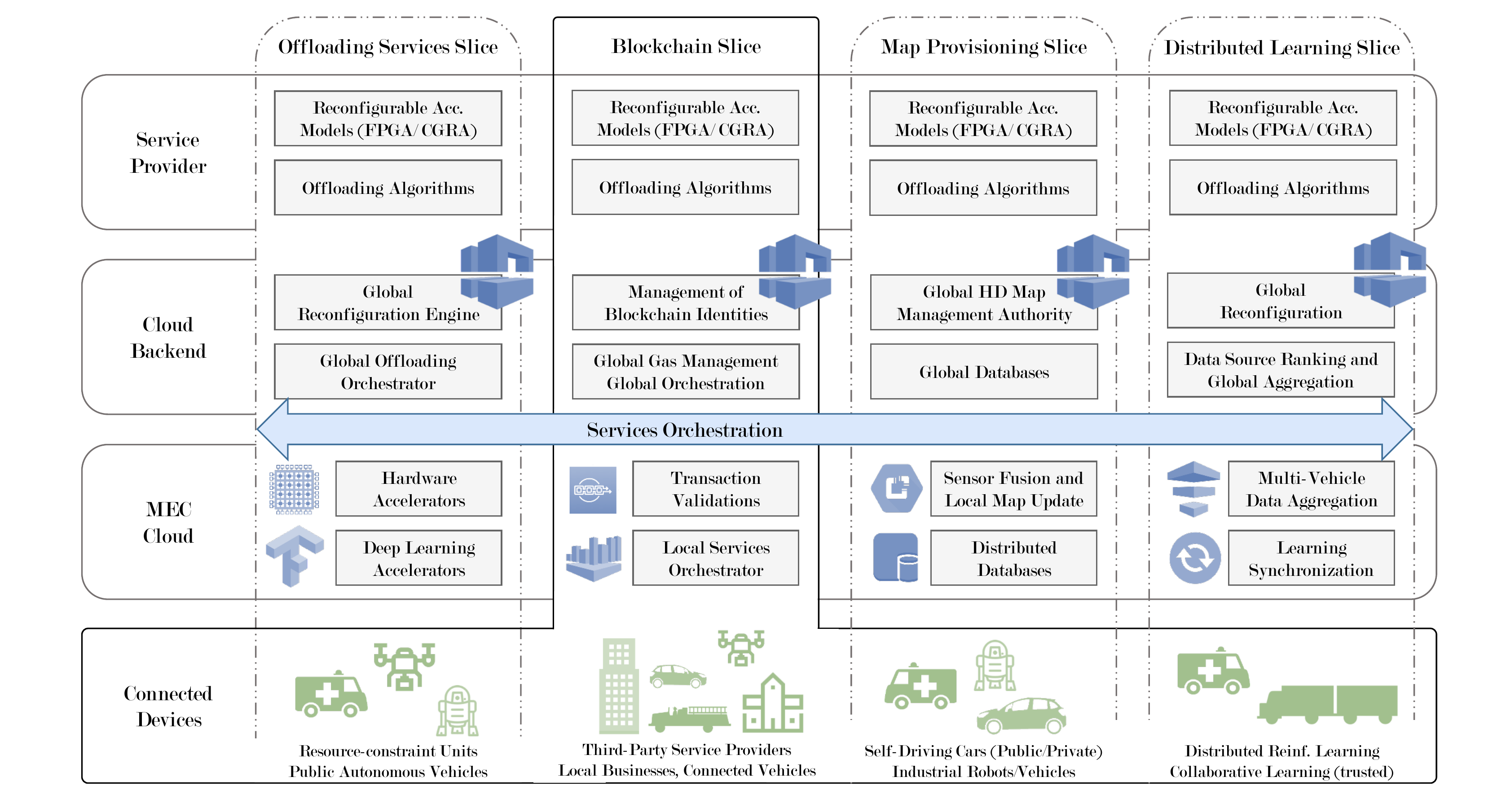}
    \caption{Architecture for Blockchain-Based MEC Autonomy Services}
    \label{fig:arch}
\end{figure*}

The integration of blockchain technology at the MEC layer has been proposed by multiple authors~\cite{xiong2018mobile, rahman2018blockchain, dai2019blockchain}. Nonetheless, these have been mostly focusing on the blockchain as a secure way of sharing or data or an immutable ledger to store transactions. However, one of the key applications of blockchains is their utilization as a robust decentralized computer that ensures the validity of execution of pieces of code called smart contracts~\cite{wood2014ethereum}. We exploit these and the consensus protocols of blockchains to provide a framework for managing edge resources and services.

\subsection{Previous works}

Xiong \textit{et al.} proposed the utilization of edge services to offer resource-constrained devices opportunities to join a blockchain by mining at the edge. Then, the end-devices share the data with third-party applications through a pricing scheme, modeling the interactions within the IoT as market activities. While their focus is on utilizing blockchains as a cryptocurrency and auditable platform, we focus on the distributed computation that smart contracts enable instead.

Liu \textit{et al.} presented a similar approach, where the MEC layer is used to offload mining operations~\cite{liu2019joint}. Nonetheless, this was part of a wider offloading framework where the focus was on deciding which offloaded operation would be cached. A similar scheme can be integrated within the offloading slice proposed in this study.

Zhu \textit{et al.}'s EdgeChain is the closest work to this work~\cite{zhu2018edgechain}. EdgeChain is a blockchain-based architecture that is utilized to place third-party applications across the MEC. We extend this idea for dynamic reconfiguration with smart contracts based on client-provider interactions, rather than considering the service providers only.

\subsection{Managing MEC with Consortium Blockchains}

A consortium blockchain such as Hyperledger deployed across the MEC layer and connected public infrastructure, which are the nodes acting as validators, can be utilized to manage the interaction between connected clients and service providers, and at the same time orchestrating the hardware resources at the MEC layer. The proposed system architecture is illustrated in Fig.~\ref{fig:arch}

We envision the existence of at least three separate network slices in order to support the aforementioned services. On one side, efficient offloading can be achieved with the on-demand reconfiguration of hardware accelerators, as well as AI accelerators. This slice focuses on low-latency in terms of fast data processing and optimization of computing resources to support as large number of connected devices as possible.

Some key benefits of this architecture are (i) identities are managed directly by the consortium blockchain and all transactions are signed and immutably recorded; (ii) the distribution of hardware resources or computing power is done through smart contracts and agreed across the MEC layer, with blockchain-enabled mobility; and (iii) a connected client, such as an autonomous robot or a self-driving car, requests a service from a third-party through the blockchain slice; if the smart contract approves the service, then the corresponding resources are configured and provisioned at the corresponding slice.

\subsection{Distributed Robotic Systems for blockchain-based services}

So far, the proposed architecture considers a distributed network architecture with distributed consensus algorithms. The last part of the piece is a distributed framework for deploying distributed robotic systems and algorithms to provide services to connected vehicles and robots. The Robot Operating System (ROS) has been the de-facto standard in production-ready robotic development for the past years~\cite{quigley2009ros}. However, wider adoption requires several challenges to be solved, including automated node discovery, real-time systems, non-ideal networks and distributed multi-robot deployments. These and other use cases are being developed within ROS2~\cite{bareis2019robots}. We believe that ROS2 will be an essential piece in connected vehicles and robots by providing a common framework and standardization to the MEC-based services. ROS2 can solve key challenges in flexible service definitions at the MEC layer. It will provide standardization of data formats, channels and deployment of algorithms, with a common underlying logic for all service providers as will be discussed in the next section.

\section{MEC for Autonomous Robotic Operation in Smart Cities}

The proposed architecture can be utilized as a general framework to provide services through MEC slicing. Nonetheless, we focus on how this architecture can be integrated to support the autonomous operation of autonomous robots and vehicles, opening the door to new applications and business opportunities in cities or areas which support such seamless integration of third-party services within the telecommunications network stack. We also outline the role of ROS2 and the blockchain in the different application scenarios.

\subsection{Provision of HD maps in real time}

Autonomous navigation in dense urban areas requires self-driving cars and more diverse autonomous robots to have the ability to localize themselves with very high-accuracy. With the current state-of-the-art, this is only possible utilizing high-definition (HD) maps of the environment~\cite{qingqing2019jdd}. However, HD maps are expensive to generate, update and maintain~\cite{hdmaps2018goldenage}. In terms of generation and real-time updates, the proposed architecture can smoothly integrate within the streaming slice a data fusion module that gathers information from different sources to obtain these maps, as described below. Regarding the maintenance, the main disadvantage of HD maps is that they require large storage on-board the vehicles, and it is impractical to keep maps of large areas within vehicles themselves. An evident solution is to provide streaming services at the MEC layer. However, this requires tight mobility and latency control. We believe that this can be achieved with the combination of ROS2 as a standardization framework, 5G and beyond networks for low-latency and predictive mechanisms to provide in advance data about the areas that robots or vehicles will travel through. ROS messages serve as a standard for data formats, which can be then processed by multiple third parties without requiring extra communication to instruct on the data structure. Moreover, the nature of ROS topics enables consistent integration within streaming services to provide HD maps, with subtopics defining various parameters, e.g. location or point cloud density.

\subsection{Online Update of Local HD Maps}

In Smart Cities, administrators can provide a framework to support the online update of HD maps and utilize existing infrastructure as a source of data. Traffic cameras and other sensors utilized for monitoring can be repurposed and their data forwarded to a data fusion scheme within a dedicated MEC slice. Moreover, connected infrastructure with enough processing power can serve to increase the range or capacity of the streaming network. The role of ROS open source algorithms is essential to deploy the state-of-the-art in multi-source and multi-sensor data fusion in public infrastructure. Moreover, the smart contracts within the blockchain can be utilized to rank the available sources of data.

By providing an open framework, city administrations open the door to new local applications such as drone delivery or various types of autonomous robots surveying, monitoring and performing other tasks across the city. This has the potential to boost both the city's economy and technology innovation.

\subsection{Distributed Reinforcement Learning}

As the robotic field has evolved over the past two decades, deep learning has become an essential aspect in complex robotics systems~\cite{sunderhauf2018limits}. In particular, reinforcement learning has allowed for traditional dynamics models to be replaced for neural networks that have been able to outperform any previous approaches. 

With the first semi-autonomous cars roaming the roads of large cities around the world, humongous amounts of data are being collected to improve the performance of deep learning algorithms. This is a process that requires offline training of neural networks. However, various distributed reinforcement learning algorithms enable robots and autonomous vehicles to have online improvements of their models not only from the real-time data and experiences but also from those of cooperating vehicles.

Offering a distributed reinforcement learning service at the MEC layer would enable connected vehicles to take advantage of the data and experiences of other vehicles to learn faster and better, with more and different experiences being analyzed in shorter periods of time. Nonetheless, such as a service would require a tight control on identities and a mechanism to ensure that model updates are valid and do provide an improvement. A permissioned blockchain provides a transparent and secure identity management framework, while the robustness and vulnerabilities in distributed multi-agent reinforcement learning is still an open problem~\cite{mhamdi2018hidden}. If raw data is provided to the learning service, then the model updates can be validated. If data is protected due to privacy concerns and only the model updates are shared, then it becomes considerably more challenging to determine the validity of a given update. A blockchain can provide part of the solution to this problem through its consensus mechanisms. They have been shown to outperform traditional consensus mechanisms in the presence of erroneous or malicious data in other scenarios within the robotics domain~\cite{strobel2018managing}.

\subsection{Offloading Services}

Reliable connectivity and existence of MEC services in a large area opens the doors to robots and vehicles relying on network-enhanced intelligence for their operation. Instead of developing and building complex robotic systems able of long-term self-supported autonomy, local organizations and businesses can build simpler products with similar capabilities, relying on computational offloading to achieve certain functionalities. Not only does this reduce the development and production cost, it also potentially decreases time-to-market, further boosting innovation.

We propose a separate slice for the offloading orchestrator and services because the focus is on optimizing computing power and reducing execution time when possible, compared to the storage and mobility requirements of the streaming slice. Even if the underlying hardware is the same, a different degree of reconfiguration is expected in order to optimize the offloading scheme.

GPU-based and FPGA-based accelerators have been widely used in sensor development and deep learning acceleration over the past few decades, being a perfect match for the requirements of edge computing~\cite{biookaghazadeh2018fpgas}. More recently, autonomous navigation, localization and mapping algorithms have started to use FPGA-based implementations for real-time matching of HD maps or odometry~\cite{qingqing2019fpga}. Some of these operations are inherently parallelizable, and therefore FPGA-based accelerators have the potential for decreasing the latency by several orders of magnitude. In the proposed architecture, we envision that dynamic reconfiguration of FPGA-based hardware accelerators will play an important role in optimizing edge resources for computational offloading, increasing the number of nodes that can be supported and reducing the execution time of different processes.

ROS services can be directly utilized in offloading schemes, where third party services simply offer these to the network. In addition, there has been a recent interest in developing ROS-compliant accelerators to match the rising computational needs~\cite{ohkawa2017rosfpgas}. Having reconfigurable hardware available on-demand at the edge can help third-party service providers integrate these solutions. The reconfiguration and provisioning of resources can then be made through smart contracts executed in the MEC-hosted blockchain. Hardware accelerator models can be naturally abstracted in terms of the number of processing units or resources required, and multiple models can co-exist within a single chip. Finally, data partitioning schemes at the blockchain level and its modular architecture with the aforementioned concepts put together an efficient, open and flexible framework for offering offloading services.

\subsection{Security Concerns}

The blockchain is a key piece in the proposed architecture as a source of trust. We consider that the main security concerns in a MEC service framework is not the exposure of data but instead its validity and reliability. This is exemplified by threats identified by ENISA such as the manipulation of the network resources orchestrator (unreliable orchestration or invalid data regarding the resource orchestration)~\cite{enisa}. As we are discussing services that support the operation of autonomous vehicles, we need to take into account that this is a safety-critical application scenario where sending wrong data to a connected vehicle or robot might put in danger pedestrians and drivers. While the blockchain is not able to provide a robust way to validate data by itself, a ranking of the different identities offering services can be created and updater in real-time. Moreover, by implementing the resource orchestration and management of edge resources with smart contracts, the reliability of services providing mission-critical data can be kept under tighter control. Finally, the immutability of the transaction record can be utilized as a posteriori to assign liability and ensure accountability.


\begin{table}
    \centering
    \caption{Characterization of MEC-based services that support the autonomous operation of connected vehicles and robots.}
    \begin{tabular}{@{}lp{1.52cm}p{1.52cm}c@{}}
        \toprule
        & \multicolumn{3}{c}{Critical Parameter} \\
                            &  \centering Latency &  \centering Throughput &  Identity/Trust \\
        \midrule
        Offloading Services  & \centering \footnotesize{\ding{52}} & \centering \footnotesize{\ding{52}} & \\[+3pt]
        Map Streaming        &   & \centering \footnotesize{\ding{52}} &   \\[+3pt]
        Distributed Learning &   &   &   \footnotesize{\ding{52}} \\
        \bottomrule
    \end{tabular}
    \label{tab:characterization}
\end{table}

\section{Discussion}

The architecture and slicing strategy defined in the previous section are based on the characterization of different services that support autonomous operation of connected robots and vehicles. This characterization is illustrated in Table~\ref{tab:characterization} from the point of view of the data and network parameters. Offloading services require low-latency and high-throughput data transmission in order to ensure full operational safety and high levels of performance. Latency is not so critical in the streaming of high-definition maps, however, if this is done such that a map of a large enough area around the vehicle is sent at a time. Finally, in distributed learning the amount of data is not critical (as raw data is not shared). Moreover, low-latency is not required because new data provided to vehicles is not critical to their operation (only improves their performance) and the learning process is long. Nonetheless, identity management and trustability are key parameters as erroneous or malicious updates to a model could have a significant impact on performance and operational safety.

The rest of this section describes the main challenges that emerge from the integration of blockchain for real-time services thar require large amount of data exchanges: scalability and storage. Nonetheless, one key point in the proposed architecture is that not all data goes through the blockchain. The purpose of the blockchain is to offer a transparent and reliable resource orchestrator, and as such service requests from connected vehicles are managed by the blockchain. However, once edge resources have been provisioned for a service and a request validated and accepted, the service is provided outside of the blockchain. Therefore, the data that supports the autonomous operation is not stored in the blockchain. In summary, it would not be very different from keeping a distributed database with all service requests.

\subsection{Scalability}

Recent advances in blockchain technology show promising results and potential for scalable and low-latency blockchain networks. Luu \textit{et al.} presented Elastico, where sharding in a permissionless blockchains was explored~\cite{luu2016secure}. Sharding is a technique that allows for distributed consensus in a network where nodes are divided in subnetworks or committees. Rather than processing and confirming all transactions globally across the network (for example through a majority consensus), each committee is in charge of processing a disjoint set of transactions, also denominated shard. In Elastico, researchers demonstrated the first sharding protocol that is secure in the presence of byzantine adversaries. Kokoris \textit{et al.} introduced OmniLedger~\cite{kokoris2018omniledger}, a decentralized and secure ledger that scales linearly with the size of the network and supports transaction confirmation times of under two seconds, potentially being able to match credit card standards in terms of transaction confirmation response time with a large enough network, compared to an average transaction confirmation time (block validation) of around ten minutes in the case of Bitcoin. While Elastico scales almost linearly with the available computation power,
OmniLedger does so with the  number of 
validators.

Regarding the scalability of Hyperledger and its channel model, the initial versions did not have a truly scalable performance. However, this has improved considerably sinve Hyperledger Fabric v1.1.0~\cite{ferris2019does}. In terms of scaling the number of channels, this has shown little performance impact with low to no degradation so far~\cite{ferris2019does}.

\subsection{Storage}

One of the key concerns when utilizing a blockchain as a service management framework is the exponential storage that will be required along time. While this is a significant factor to take into account with early networks such as Bitcoin or Ethereum, next-generation blockchains have tackled this issue. Omniledger introduced state-blocks decreasing storage costs~\cite{kokoris2018omniledger}. More recently, RapidChain achieved a storage saving factor of over 5x when compared to Omniledger, and 16x when compared with Elastico and Bitcoin-like blockchains~\cite{zamani2018rapidchain}. In any case, the storage needs of a blockchain and its impact on performance can be controlled by truncating it or defining a fixed lifecycle. While this is a challenge in permissionless open blockchains, a strategy can be defined by public authorities for a permissioned blockchain such as Hyperledger if the infrastructure being used to validate transactions depends on the same public authorities.
\section{Conclusion}

We have presented a system architecture for offering autonomy support services at the MEC layer in 5G and beyond networks. Our architecture combines network slicing with blockchain technology and distributed robotic systems. In particular, we envision a future of connected vehicles and robots where higher degrees of autonomy will be achieved through third-party services at the MEC layer. These services and the provisioning of MEC resources can be managed with permissioned blockchain networks such as Hyperledger, where offloading orchestrators and streaming orchestrators are implemented through smart contracts.

In future work, we will utilize ROS2 and Hyperledger to provide a proof-of-concept for the proposed architecture, and the bottlenecks and limitations will be analyzed.


\section*{Acknowledgements}
This work was supported by the Academy of Finland's AutoSOS project with grant number 328755, the NSFC grant number 61876039, and the Shanghai Platform for Neuromorphic and AI Chips (NeuHeilium).



\bibliographystyle{unsrt}
\bibliography{ref}

\begin{thebibliography}{10}

\bibitem{dohler2017internet}
Mischa Dohler, Toktam Mahmoodi, Maria~A Lema, Massimo Condoluci, Fragkiskos
  Sardis, Konstantinos Antonakoglou, and Hamid Aghvami.
\newblock Internet of skills, where robotics meets ai, 5g and the tactile
  internet.
\newblock In {\em 2017 European Conference on Networks and Communications
  (EuCNC)}, pages 1--5. IEEE, 2017.

\bibitem{li2017intelligent}
Rongpeng Li, Zhifeng Zhao, Xuan Zhou, Guoru Ding, Yan Chen, Zhongyao Wang, and
  Honggang Zhang.
\newblock Intelligent 5g: When cellular networks meet artificial intelligence.
\newblock {\em IEEE Wireless communications}, 24(5):175--183, 2017.

\bibitem{palattella2016internet}
Maria~Rita Palattella, Mischa Dohler, Alfredo Grieco, Gianluca Rizzo, Johan
  Torsner, Thomas Engel, and Latif Ladid.
\newblock Internet of things in the 5g era: Enablers, architecture, and
  business models.
\newblock {\em IEEE Journal on Selected Areas in Communications},
  34(3):510--527, 2016.

\bibitem{backman2017blockchain}
Jere Backman, Seppo Yrj{\"o}l{\"a}, Kristiina Valtanen, and Olli
  M{\"a}mmel{\"a}.
\newblock Blockchain network slice broker in 5g: Slice leasing in factory of
  the future use case.
\newblock In {\em 2017 Internet of Things Business Models, Users, and
  Networks}, pages 1--8. IEEE, 2017.

\bibitem{voigtlander20175g}
F.~Voigtl{\"a}nder \textit{et al.}
\newblock 5g for robotics: Ultra-low latency control of distributed robotic
  systems.
\newblock In {\em 2017 International Symposium on Computer Science and
  Intelligent Controls (ISCSIC)}, pages 69--72. IEEE, 2017.

\bibitem{queralta2020blockchain}
Jorge~Pena Queralta and Tomi Westerlund.
\newblock Blockchain-powered collaboration in heterogeneous swarms of robots.
\newblock {\em Frontiers in Robotics and AI}, 2020.

\bibitem{wang2019edge}
Xiaofei Wang, Yiwen Han, Chenyang Wang, Qiyang Zhao, Xu~Chen, and Min Chen.
\newblock In-edge ai: Intelligentizing mobile edge computing, caching and
  communication by federated learning.
\newblock {\em IEEE Network}, 33(5):156--165, 2019.

\bibitem{addad2018towards}
Rami~Akrem Addad, Tarik Taleb, Miloud Bagaa, Diego Leonel~Cadette Dutra, and
  Hannu Flinck.
\newblock Towards modeling cross-domain network slices for 5g.
\newblock In {\em 2018 IEEE Global Communications Conference (GLOBECOM)}, pages
  1--7. IEEE, 2018.

\bibitem{husain2018mobile}
Syed Husain, Andreas Kunz, Athul Prasad, Konstantinos Samdanis, and JaeSeung
  Song.
\newblock Mobile edge computing with network resource slicing for
  internet-of-things.
\newblock In {\em 2018 IEEE 4th World Forum on Internet of Things (WF-IoT)},
  pages 1--6. IEEE, 2018.

\bibitem{quigley2009ros}
Morgan Quigley, Ken Conley, Brian Gerkey, Josh Faust, Tully Foote, Jeremy
  Leibs, Rob Wheeler, and Andrew~Y Ng.
\newblock Ros: an open-source robot operating system.
\newblock In {\em ICRA workshop on open source software}, volume~3, page~5.
  Kobe, Japan, 2009.

\bibitem{sabella2016mobile}
Dario Sabella, Alessandro Vaillant, Pekka Kuure, Uwe Rauschenbach, and Fabio
  Giust.
\newblock Mobile-edge computing architecture: The role of mec in the internet
  of things.
\newblock {\em IEEE Consumer Electronics Magazine}, 5(4):84--91, 2016.

\bibitem{qingqing2019monocular}
Li~Qingqing, Jorge~Pena Queralta, Tuan~Nguyen Gia, and Tomi Westerlund.
\newblock Offloading monocular visual odometry with edge computing: Optimizing
  image quality in multi-robot systems.
\newblock In {\em Proceedings of the 2019 5th International Conference on
  Systems, Control and Communications}, pages 22--26, 2019.

\bibitem{zhang2016energy}
Ke~Zhang, Yuming Mao, Supeng Leng, Quanxin Zhao, Longjiang Li, Xin Peng,
  Li~Pan, Sabita Maharjan, and Yan Zhang.
\newblock Energy-efficient offloading for mobile edge computing in 5g
  heterogeneous networks.
\newblock {\em IEEE access}, 4:5896--5907, 2016.

\bibitem{zhang2017mobile}
Ke~Zhang, Yuming Mao, Supeng Leng, Yejun He, and Yan Zhang.
\newblock Mobile-edge computing for vehicular networks: A promising network
  paradigm with predictive off-loading.
\newblock {\em IEEE Vehicular Technology Magazine}, 12(2):36--44, 2017.

\bibitem{bareis2019robots}
Agata Barei{\'s}, Micha{\l} Barei{\'s}, and Christian Bettstetter.
\newblock Robots that sync and swarm: a proof of concept in ros 2.
\newblock In {\em 2019 International Symposium on Multi-Robot and Multi-Agent
  Systems (MRS)}, pages 98--104. IEEE, 2019.

\bibitem{wood2014ethereum}
Gavin Wood et~al.
\newblock Ethereum: A secure decentralised generalised transaction ledger.
\newblock {\em Ethereum project yellow paper}, 151(2014):1--32, 2014.

\bibitem{seif2016autonomous}
Heiko~G Seif and Xiaolong Hu.
\newblock Autonomous driving in the icity—hd maps as a key challenge of the
  automotive industry.
\newblock {\em Engineering}, 2(2):159--162, 2016.

\bibitem{hong2019rules}
Joey Hong, Benjamin Sapp, and James Philbin.
\newblock Rules of the road: Predicting driving behavior with a convolutional
  model of semantic interactions.
\newblock In {\em Proceedings of the IEEE Conference on Computer Vision and
  Pattern Recognition}, pages 8454--8462, 2019.

\bibitem{floreano2015science}
Dario Floreano and Robert~J Wood.
\newblock Science, technology and the future of small autonomous drones.
\newblock {\em Nature}, 521(7553):460--466, 2015.

\bibitem{xiong2018mobile}
Zehui Xiong, Yang Zhang, Dusit Niyato, Ping Wang, and Zhu Han.
\newblock When mobile blockchain meets edge computing.
\newblock {\em IEEE Communications Magazine}, 56(8):33--39, 2018.

\bibitem{rahman2018blockchain}
MD~Abdur Rahman, M~Shamim Hossain, George Loukas, Elham Hassanain, Syed~Sadiqur
  Rahman, Mohammed~F Alhamid, and Mohsen Guizani.
\newblock Blockchain-based mobile edge computing framework for secure therapy
  applications.
\newblock {\em IEEE Access}, 6:72469--72478, 2018.

\bibitem{dai2019blockchain}
Yueyue Dai, Du~Xu, Sabita Maharjan, Zhuang Chen, Qian He, and Yan Zhang.
\newblock Blockchain and deep reinforcement learning empowered intelligent 5g
  beyond.
\newblock {\em IEEE Network}, 33(3):10--17, 2019.

\bibitem{ferrer2018blockchain}
EC~Ferrer.
\newblock The blockchain: a new framework for robotic swarm systems.
\newblock In {\em Proceedings of the Future Technologies Conference}, pages
  1037--1058. Springer, 2018.

\bibitem{strobel2018managing}
Volker Strobel, Eduardo Castell{\'o}~Ferrer, and Marco Dorigo.
\newblock Managing byzantine robots via blockchain technology in a swarm
  robotics collective decision making scenario.
\newblock {\em AAMAS, Springer}, 2018.

\bibitem{smart_city_applications}
Kehua Su, Jie Li, and Hongbo Fu.
\newblock Smart city and the applications.
\newblock In {\em 2011 international conference on electronics, communications
  and control (ICECC)}, pages 1028--1031. IEEE, 2011.

\bibitem{queralta2019collaborative}
Jorge~Pena Queralta, Tuan~Nguyen Gia, Hannu Tenhunen, and Tomi Westerlund.
\newblock Collaborative mapping with ioe-based heterogeneous vehicles for
  enhanced situational awareness.
\newblock In {\em 2019 IEEE Sensors Applications Symposium (SAS)}, pages 1--6.
  IEEE, 2019.

\bibitem{concptualizing_smart_cities}
Taewoo Nam and Theresa~A Pardo.
\newblock Conceptualizing smart city with dimensions of technology, people, and
  institutions.
\newblock In {\em Proceedings of the 12th annual international digital
  government research conference: digital government innovation in challenging
  times}, pages 282--291, 2011.

\bibitem{taleb2017multi}
Tarik Taleb, Konstantinos Samdanis, Badr Mada, Hannu Flinck, Sunny Dutta, and
  Dario Sabella.
\newblock On multi-access edge computing: A survey of the emerging 5g network
  edge cloud architecture and orchestration.
\newblock {\em IEEE Communications Surveys \& Tutorials}, 19(3):1657--1681,
  2017.

\bibitem{campolo2017slicing}
Claudia Campolo, Antonella Molinaro, Antonio Iera, and Francesco Menichella.
\newblock 5g network slicing for vehicle-to-everything services.
\newblock {\em IEEE Wireless Communications}, 24(6):38--45, 2017.

\bibitem{mei2019intelligent}
Jie Mei, Xianbin Wang, and Kan Zheng.
\newblock Intelligent network slicing for v2x services toward 5g.
\newblock {\em IEEE Network}, 33(6):196--204, 2019.

\bibitem{etsi2018mec}
S~Kekki \textit{et al.}
\newblock Mec in 5g networks.
\newblock {\em ETSI white paper}, 28:1--28, 2018.

\bibitem{hu2015mobile}
Yun~Chao Hu, Milan Patel, Dario Sabella, Nurit Sprecher, and Valerie Young.
\newblock Mobile edge computing—a key technology towards 5g.
\newblock {\em ETSI white paper}, 11(11):1--16, 2015.

\bibitem{nextgen2016study}
3GPP.
\newblock Study on architecture for next-generation system rel. 14.
\newblock {\em Techical Report}, 23.799, 2016.

\bibitem{alliance2016description}
N.~Alliance.
\newblock Description of network slicing concept.
\newblock {\em NGMN 5G P}, 1:1, 2016.

\bibitem{giust2018multi}
Fabio Giust, Vincenzo Sciancalepore, Dario Sabella, Miltiades~C Filippou,
  Simone Mangiante, Walter Featherstone, and Daniele Munaretto.
\newblock Multi-access edge computing: The driver behind the wheel of
  5g-connected cars.
\newblock {\em IEEE Communications Standards Magazine}, 2(3):66--73, 2018.

\bibitem{enisa}
The European Union~Agency for Cybersecurity.
\newblock Threat assessment for the fifth generation of mobile
  telecommunications networks (5g).
\newblock {\em ENISA THREAT LANDSCAPE FOR 5G NETWORKS}, 2019.

\bibitem{vukolic2017rethinking}
M.~Vukoli{\'c}.
\newblock Rethinking permissioned blockchains.
\newblock In {\em Proceedings of the ACM Workshop on Blockchain,
  Cryptocurrencies and Contracts}, pages 3--7. ACM, 2017.

\bibitem{zheng2017overview}
Zibin Zheng, Shaoan Xie, Hongning Dai, Xiangping Chen, and Huaimin Wang.
\newblock An overview of blockchain technology: Architecture, consensus, and
  future trends.
\newblock In {\em 2017 IEEE international congress on big data (BigData
  congress)}, pages 557--564. IEEE, 2017.

\bibitem{cachin2016architecture}
C.~Cachin.
\newblock Architecture of the hyperledger blockchain fabric.
\newblock In {\em Workshop on distributed cryptocurrencies and consensus
  ledgers}, volume 310, page~4, 2016.

\bibitem{androulaki2018hyperledger}
Elli Androulaki, Artem Barger, Vita Bortnikov, Christian Cachin, Konstantinos
  Christidis, Angelo De~Caro, David Enyeart, Christopher Ferris, Gennady
  Laventman, Yacov Manevich, et~al.
\newblock Hyperledger fabric: a distributed operating system for permissioned
  blockchains.
\newblock In {\em Proceedings of the thirteenth EuroSys conference}, pages
  1--15, 2018.

\bibitem{ibm2018hyperledger}
Sharon Cocco and G~Singh.
\newblock Top 6 technical advantages of hyperledger fabric for blockchain
  networks, 2018.

\bibitem{liu2019joint}
Mengting Liu, F~Richard Yu, Yinglei Teng, Victor~CM Leung, and Mei Song.
\newblock Joint computation offloading and content caching for wireless
  blockchain networks.
\newblock In {\em IEEE INFOCOM 2018-IEEE Conference on Computer Communications
  Workshops (INFOCOM WKSHPS)}, pages 517--522. IEEE, 2018.

\bibitem{zhu2018edgechain}
He~Zhu, Changcheng Huang, and Jiayu Zhou.
\newblock Edgechain: Blockchain-based multi-vendor mobile edge application
  placement.
\newblock In {\em 2018 4th IEEE Conference on Network Softwarization and
  Workshops (NetSoft)}, pages 222--226. IEEE, 2018.

\bibitem{qingqing2019jdd}
Li~Qingqing, Jorge~Pena Queralta, Tuan~Nguyen Gia, Zhuo Zou, and Tomi
  Westerlund.
\newblock Multi sensor fusion for navigation and mapping in autonomous
  vehicles: Accurate localization in urban environments.
\newblock {\em The 9th IEEE CIS-RAM}, 2019.

\bibitem{hdmaps2018goldenage}
Synced.
\newblock The golden age of hd mapping for autonomous driving.
\newblock {\em Medium}, 2018.

\bibitem{sunderhauf2018limits}
Niko S{\"u}nderhauf, Oliver Brock, Walter Scheirer, Raia Hadsell, Dieter Fox,
  J{\"u}rgen Leitner, Ben Upcroft, Pieter Abbeel, Wolfram Burgard, Michael
  Milford, et~al.
\newblock The limits and potentials of deep learning for robotics.
\newblock {\em The International Journal of Robotics Research},
  37(4-5):405--420, 2018.

\bibitem{mhamdi2018hidden}
El~Mahdi~El Mhamdi, Rachid Guerraoui, and S{\'e}bastien Rouault.
\newblock The hidden vulnerability of distributed learning in byzantium.
\newblock {\em arXiv preprint arXiv:1802.07927}, 2018.

\bibitem{biookaghazadeh2018fpgas}
Saman Biookaghazadeh, Ming Zhao, and Fengbo Ren.
\newblock Are fpgas suitable for edge computing?
\newblock In {\em $\{$USENIX$\}$ Workshop on Hot Topics in Edge Computing
  (HotEdge 18)}, 2018.

\bibitem{qingqing2019fpga}
L~Qingqing, F~Yuhong, J~Pena Queralta, Tuan~Nguyen Gia, Hannu Tenhunen, Zhuo
  Zou, and Tomi Westerlund.
\newblock Edge computing for mobile robots: multi-robot feature-based lidar
  odometry with fpgas.
\newblock In {\em 2019 Twelfth International Conference on Mobile Computing and
  Ubiquitous Network (ICMU)}, pages 1--2. IEEE, 2019.

\bibitem{ohkawa2017rosfpgas}
T~Ohkawa, Y~Ishida, Y~Sugata, and H~Tamukoh.
\newblock Ros-compliant fpga component technology—fpga installation into ros,
  2017.

\bibitem{luu2016secure}
Loi Luu, Viswesh Narayanan, Chaodong Zheng, Kunal Baweja, Seth Gilbert, and
  Prateek Saxena.
\newblock A secure sharding protocol for open blockchains.
\newblock In {\em Proceedings of the 2016 ACM SIGSAC Conference on Computer and
  Communications Security}, pages 17--30, 2016.

\bibitem{kokoris2018omniledger}
Eleftherios Kokoris-Kogias, Philipp Jovanovic, Linus Gasser, Nicolas Gailly,
  Ewa Syta, and Bryan Ford.
\newblock Omniledger: A secure, scale-out, decentralized ledger via sharding.
\newblock In {\em 2018 IEEE Symposium on Security and Privacy (SP)}, pages
  583--598. IEEE, 2018.

\bibitem{ferris2019does}
Christopher Ferris.
\newblock “does hyperledger fabric perform at scale?
\newblock {\em Blockchain Pulse: IBM Blockchain Blog}, 2, 2019.

\bibitem{zamani2018rapidchain}
Mahdi Zamani, Mahnush Movahedi, and Mariana Raykova.
\newblock Rapidchain: Scaling blockchain via full sharding.
\newblock In {\em Proceedings of the 2018 ACM SIGSAC Conference on Computer and
  Communications Security}, pages 931--948, 2018.

\end{thebibliography}

\end{document}